\documentclass{aa}

\begin{document}

\thesaurus{07(08.03.4;08.05.2;08.09.2 A\,0535+26;08.02.1;08.14.1;13.25.5)} 

\title{Detection of X-ray pulsations from the Be/X-ray transient 
A\,0535+26 during a disc loss phase of the primary}

\author{I.~Negueruela\inst{1}
\and P.~Reig\inst{2,3} 
\and M.~H.~Finger\inst{4}
\and P. Roche\inst{5}}                   
                                                            
\institute{SAX SDC, ASI, c/o Nuova Telespazio, via Corcolle 19, I00131
Rome, Italy
\and Foundation for Research and Technology-Hellas, 711 10, Heraklion,
Crete, Greece
\and Physics Department, University of Crete, 710 03, Heraklion, Crete, Greece
\and Space Science Laboratory, ES84, NASA/Marshall Space Flight
Center, Huntsville, AL 35812, USA
\and Department of Physics \& Astronomy, University of Leicester, Leicester,
LE1 7RH, U.K.}

\mail{ignacio@tocai.sdc.asi.it}

\date{Received    / Accepted     }

\titlerunning{Pulsations from A\,0535+26}
\authorrunning{Negueruela et al.}
\maketitle 

\begin{abstract}
Using the {\em RossiXTE} experiment, we detect weak X-ray emission from the 
recurrent Be/X-ray transient A\,0535+26 at a time when the optical counterpart
V725 Tau displayed H$\alpha$ in absorption, indicating the absence of a 
circumstellar disc. The X-ray radiation is strongly modulated at the 
103.5-s pulse
period of the neutron star, confirming that it originates from A\,0535+26. 
The source is weaker than in previous quiescence detections by two orders of
magnitude and should be in the centrifugal inhibition regime. We show that 
the X-ray luminosity cannot be due to accretion on to the magnetosphere of
the neutron star. Therefore this detection represents a new state of the
accreting pulsar. We speculate that the X-ray emission can be due to some 
matter leaking through the magnetospheric barrier or thermal 
radiation from the neutron star surface due to crustal heating. The
observed luminosity is probably compatible with recent predictions of
thermal radiation from X-ray transients in quiescence.
The detection of the X-ray source in the inhibition regime implies a 
reduced density in the outflow from the Be companion during its
disc-less phase.
 \end{abstract}


\section{Introduction}

Be/X-ray binaries are X-ray sources composed of a Be star and a neutron star.
Most of these systems are transient X-ray pulsars displaying strong
outbursts in which their X-ray luminosity increases by a factor $\ga 100$
(see Negueruela 1998).
In addition, those systems in which the neutron star does not rotate fast
enough for centrifugal inhibition of accretion to be effective (see Stella
et al. 1986) display persistent X-ray luminosity at a level $\la 10^{35}$ erg
s$^{-1}$.
The high-energy radiation is believed to arise due to accretion of material 
associated with the Be star by the compact object. It has long been known
that accretion from the fast polar wind that is detected in the UV resonance
lines of the Be primaries cannot provide the observed luminosities, even for 
detections at the weakest level (see Waters et al. 1988 and references 
therein; see also the calculations made for X Persei
by Telting et al. 1998). Therefore it is believed that
the material accreted comes from the dense equatorial disc that surrounds 
the Be star. Waters et al. (1989) modelled the radial outflow as a relatively
slow ($\sim 100$ km s$^{-1}$) dense wind. However most modern models for Be
stars consider much slower outflows, due to strong evidence for rotationally 
dominated motion (quasi-Keplerian discs). This is due not only to the
line shapes (see Hanuschik et al. 1996), which set an upper limit on 
the bulk motion at $v \la 3\:{\rm km}\,{\rm s}^{-1}$ (Hanuschik 2000), but 
also to the success of the Global One-Armed Oscillation model (which
can only work in quasi-Keplerian discs) at explaining V/R variability
in Be stars (Hummel \& Hanuschik 1997; Okazaki 1997ab). 
The viscous decretion disc model (Okazaki 1997b; Porter 1999; 
Negueruela \& Okazaki 2000) considers
material in  quasi-Keplerian orbit with an initially very subsonic outflow 
velocity that is gradually accelerated by gas pressure and
becomes supersonic at distances $\sim 100 R_{*}$, i.e., much further than the
orbits of neutron stars in Be/X-ray transients.

The transient A\,0535+26 is one of the best studied Be/X-ray binaries 
(Clark et al. 1998 and references therein). It contains a slowly 
rotating ($P_{{\rm s}} = 103 \: 
{\rm s}$) neutron star in a relatively wide ($P_{{\rm orb}} = 
110.3 \: {\rm d}$) and eccentric ($e = 0.47$) 
orbit around the B0IIIe star V725 Tau (see Finger et al. 1996; Steele et
al. 1998). After its last giant outburst in February 1994 (Clark et al. 
1998; Negueruela et al. 1998), the strength of the emission lines in 
the optical spectrum of V725 Tau has declined steadily. 
The last normal (periodic) outburst 
took place in September 1994 and the source has since not been detected 
by the BATSE experiment on board {\em CGRO}.

\section{Observations}

\subsection{Optical spectroscopy}

\begin{figure}
\begin{picture}(250,250)
\put(0,0){\includegraphics{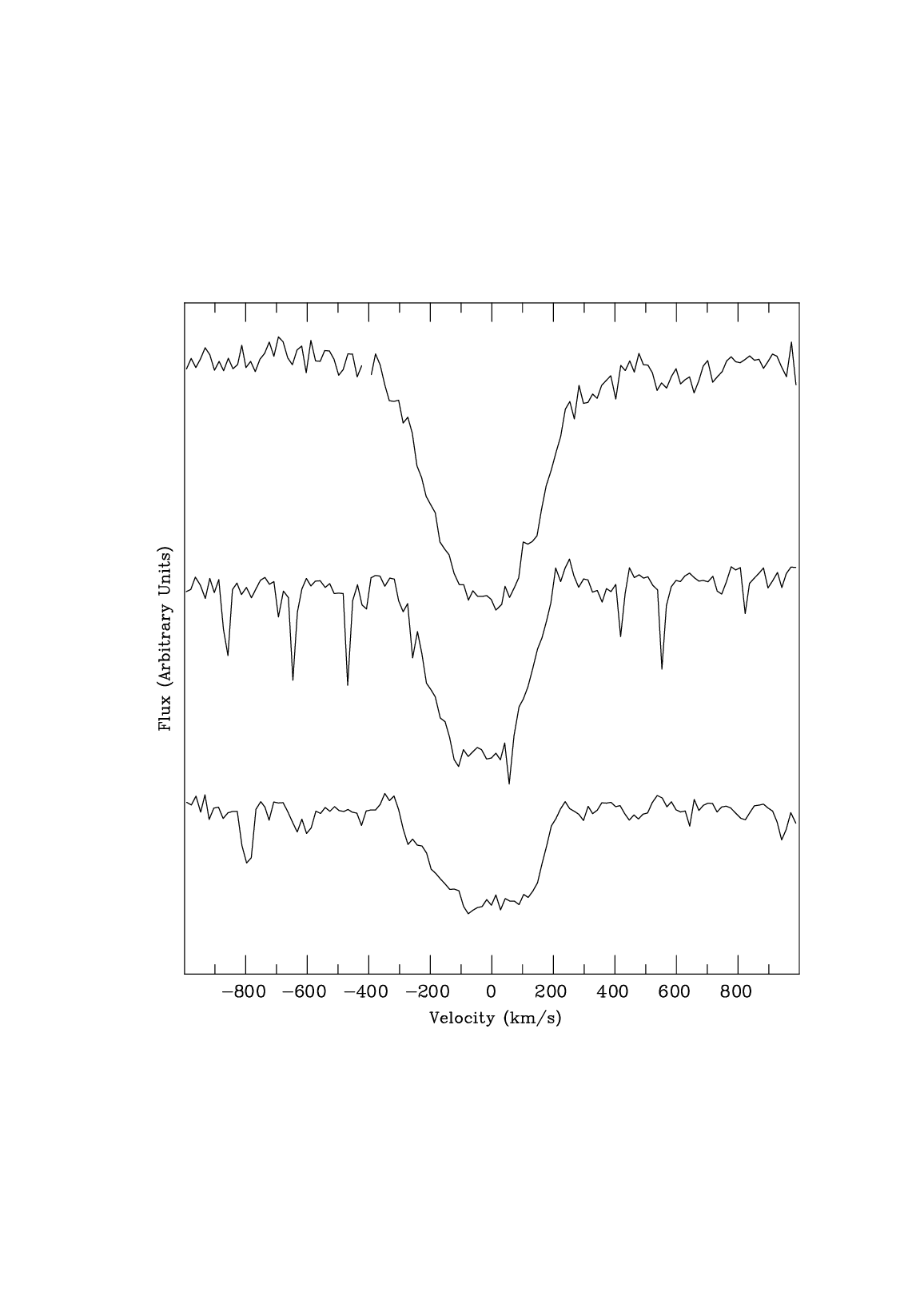}}
\end{picture}
\caption{Spectroscopy of V725 Tau taken on November 7, 1998, with the WHT
and UES. The lines are, from top to bottom H$\beta$, H$\alpha$ and \ion{He}{i}
$\lambda$6678\AA. The spectra have been divided by a spline fit to the
continuum and arbitrarily offset for display. The sharp lines surrounding
H$\alpha$ are atmospheric water vapour features.} 
\label{fig:opt} 
\end{figure}

V725 Tau, the optical counterpart to A\,0535+26, was observed on November 7th
1998, using the 4.2-m William Herschel Telescope, 
located at the Observatorio del Roque de los Muchachos, La Palma, Spain.
The telescope was equipped with the Utrecht Echelle Spectrograph using the
31.6 lines/mm echelle centred at H$\alpha$ and the SITe1 CCD camera. This
configuration gives a resolution $R \sim 40000$ over the range $\sim 4600 
\,-\, 10200$ \AA. The data have been reduced using the {\em Starlink} packages
{\sc ccdpack} (Draper 1998), {\sc echomop} (Mills et al. 1997) and {\sc dipso}
(Howarth et al. 1997). A detailed analysis of the whole spectrum is left for a 
forthcoming paper. In Figure~\ref{fig:opt}, we show the shape of H$\alpha$,
H$\beta$ and \ion{He}{i} $\lambda$6678\AA. When in emission, these three
lines sample most of the radial extent of the circumstellar envelope. However,
it is apparent that the lines seen in Fig.~\ref{fig:opt} correspond to 
photospheric absorption from the underlying star. The emission 
contribution from circumstellar material, if any, is certainly very small.
The asymmetry in the shape of H$\alpha$ and \ion{He}{i} $\lambda$6678\AA\
suggests that some fast-moving material is present close to the 
stellar surface (see Hanuschik et al. 1993; 
Rivinius et al. 1998 for the discussion of low-level activity
in disc-less Be stars), but 
the circumstellar disc is basically absent. H$\beta$, which, when in emission,
is typically produced at distances of a few $R_{*}$, looks completely 
photospheric. In Be stars H$\alpha$ probes a region extending to 
$\sim 10\: R_{*}$, as measured from line-peak separation (Hummel \&
Vrancken 1995) and direct imaging (Quirrenbach et al. 1997). Again, 
circumstellar material seems to be almost absent from this region. There
is weak emission emission in-filling at the line centre -- transient emission 
components have been seen in this star during the disc-less state
(Haigh et al. 2000), a behaviour typical of disc-less Be stars 
(Rivinius et al. 1998 and references therein).

\subsection{X-ray observations}

Observations of the source were taken using the Proportional Counter Array 
(PCA) on board {\em RossiXTE} on 1998 August 21 and 1998 November 12 for a 
total on-source time of 4170 s  and 2250 s, respectively.  

In both observations there is an excess of $\sim4$ counts/s/(5 PCU) in the
2.5\,--15 keV range of the Standard2 data compared to the faint source
background model. Fits to power-law models with interstellar absorption
result in flux estimates of $6\times10^{-12}$ and 
$9\times10^{-12}\:{\rm erg}\,{\rm cm}^{-2}\,{\rm s}^{-1}$ (2\,--\,10 keV) for
the two observations respectively. These can only be considered as upper
limits on the flux, due to the uncertain contribution of diffuse Galactic 
Disc emission to the count rates (Valinia \& Marshall 1998).

\subsubsection{Timing analysis}

The issue of whether the source of high energy radiation was active or not
during the low activity optical phase can be solved by searching for the
previously reported X-ray pulsations at $\approx$103.5-s spin period
(e.g., Finger et al. 1996). In order to improve the signal-to-noise, 
we accumulated events from the top anode layer of the detectors. 
We also used the latest version of 
the faint background model. 

For the power spectral analysis we selected a stretch of continuous data
and divided it into intervals of 309 s. A power spectrum was
obtained for each interval and the results averaged together. 
Given that the pulse frequency ($\nu \approx 0.0097$ Hz)
lies on a region dominated by red noise, we have to correct for such noise
if the statistical significance of the pulsations are to be established.  
First, we fitted the Poisson level by restricting ourselves to the
frequency range 0.2\,--\,0.4 Hz, that is far away from the region where the
red noise component may contribute appreciably. The strongest peak in 
the power spectrum corresponds to $\approx 103.5\:{\mathrm s}$.

\begin{figure}
\begin{picture}(250,280)
\put(0,0){\includegraphics{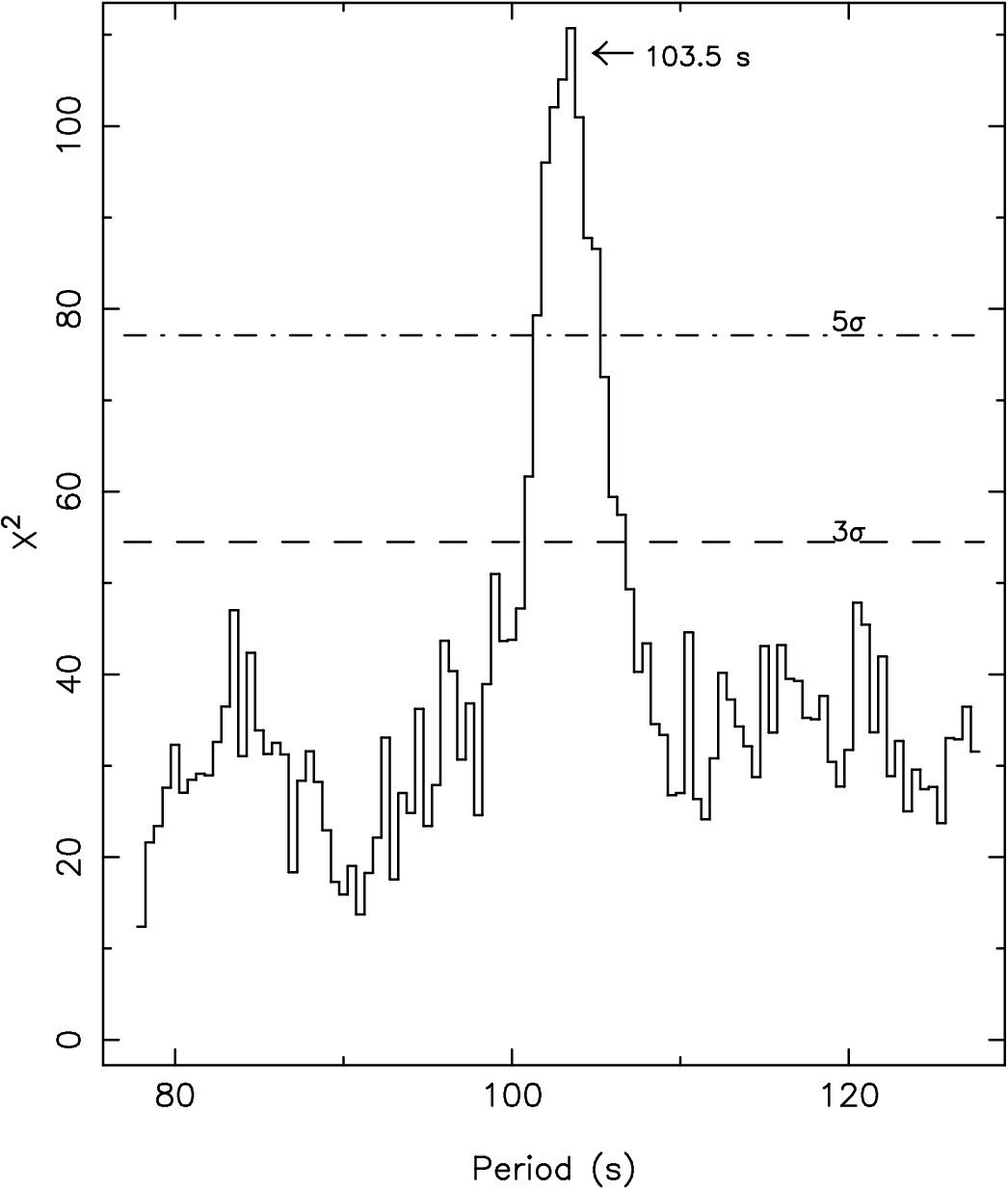}}
\end{picture}
\caption{3\,--\,20 keV light-curve folded over a range of trial periods 
(epoch folding) for the November 1998 observation. The arrival times 
were corrected to the
solar barycentre. The dashed and dash-dotted lines represent the 3-$\sigma$
and 5-$\sigma$ detection levels considering that all trial periods have the
same probability.} 
\label{fig:epoch} 
\end{figure}

We also searched
for periodicities in the light curves by folding the data over a period
range and determining the $\chi^{2}$ of the folded light-curve (epoch-folding 
technique).  In this case we used 20 phase bins (19 degrees of
freedom) and a range of 100 periods, around the expected period.  This
method has the advantage that the result is not affected by the presence
of gaps in the data, hence a longer baseline can be considered than
with Fourier analysis. Times
in the background subtracted light-curve were converted into times
at the solar-system barycentre.  The result for the 1998 November
observation is shown in Fig~\ref{fig:epoch}. We found that the peak at 
$\sim 103.5\:{\mathrm s}$
is significant at $> 5 \sigma$, confirming that the source was active
during the observations.

It is
worth mentioning that the detection levels shown in Fig.~\ref{fig:epoch}
were obtained without {\em a priori} knowledge of the frequency of pulsations. 
In other words, we searched for pulsations in the frequency/period range
shown in the figure.  If we take into account the fact that we are 
interested in the pulse period {\em at} $103.5\:{\mathrm s}$, 
the peak becomes still more significant. 

\begin{figure}
\begin{picture}(250,280)
\put(0,0){\includegraphics{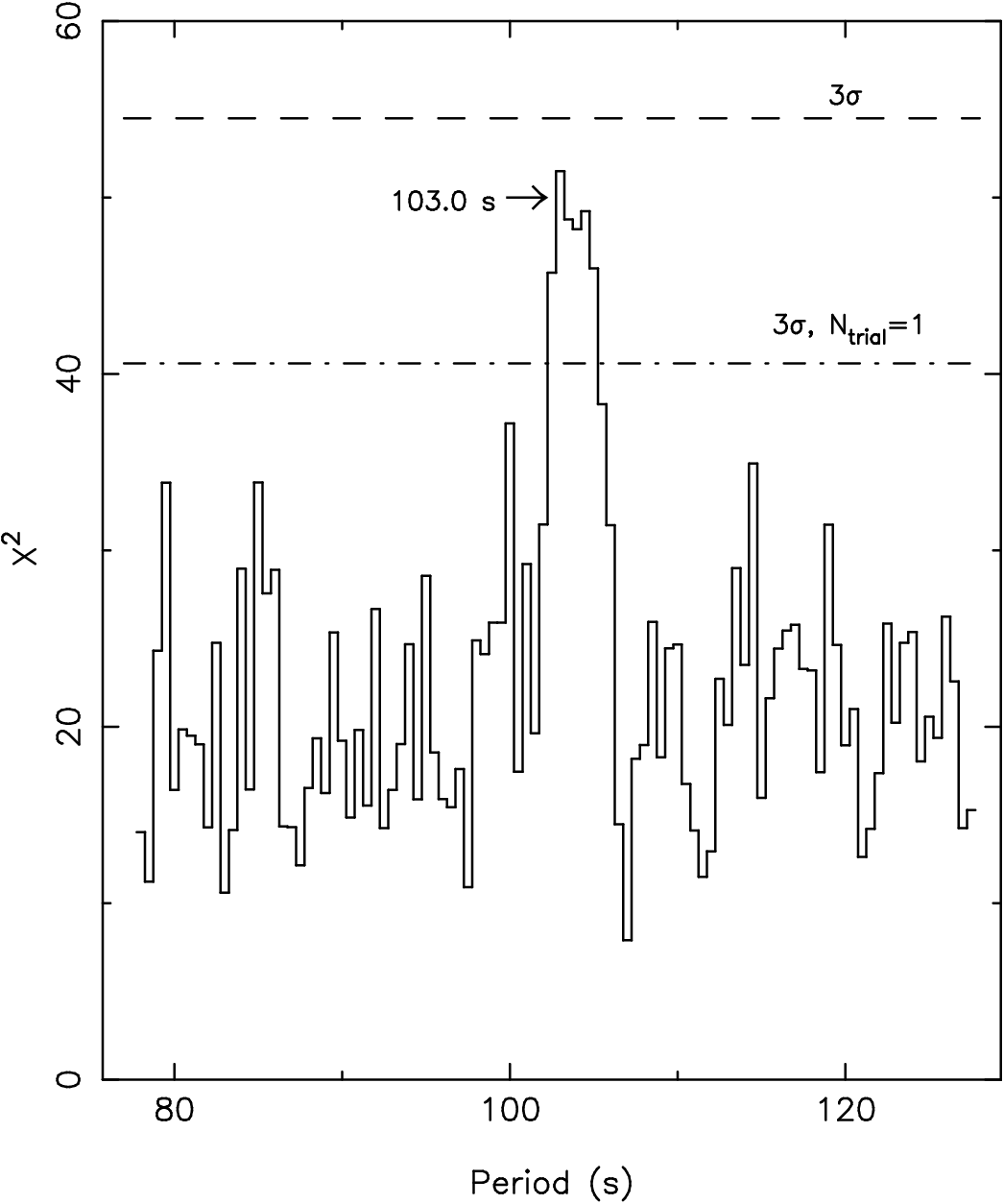}}
\end{picture}
\caption{Same as Fig.~\ref{fig:epoch} but for the August 1998 observation. 
The dash-dotted line
represents the 3-$\sigma$ detection level and was obtained by considering
one trial period only (that at which pulsations are expected).} 
\label{fig:epoch2} 
\end{figure}

The analysis of the 1998 August observation provides a much less significant
detection. A peak at the expected frequency ($\nu \approx 
0.0097$ Hz) is seen in the power spectrum. However, epoch-folding 
analysis gave a significance of $3\sigma$ only when we considered one 
single period, that is, the number of trials is one. A search for 
pulsations in a period range 
did not yield any maximum above the 3-$\sigma$ detection level although a peak 
at the 103-s period is present (see Fig. \ref{fig:epoch2})

\subsubsection{Pulse shape}

\begin{figure}
\begin{picture}(250,280)
\put(0,0){\includegraphics{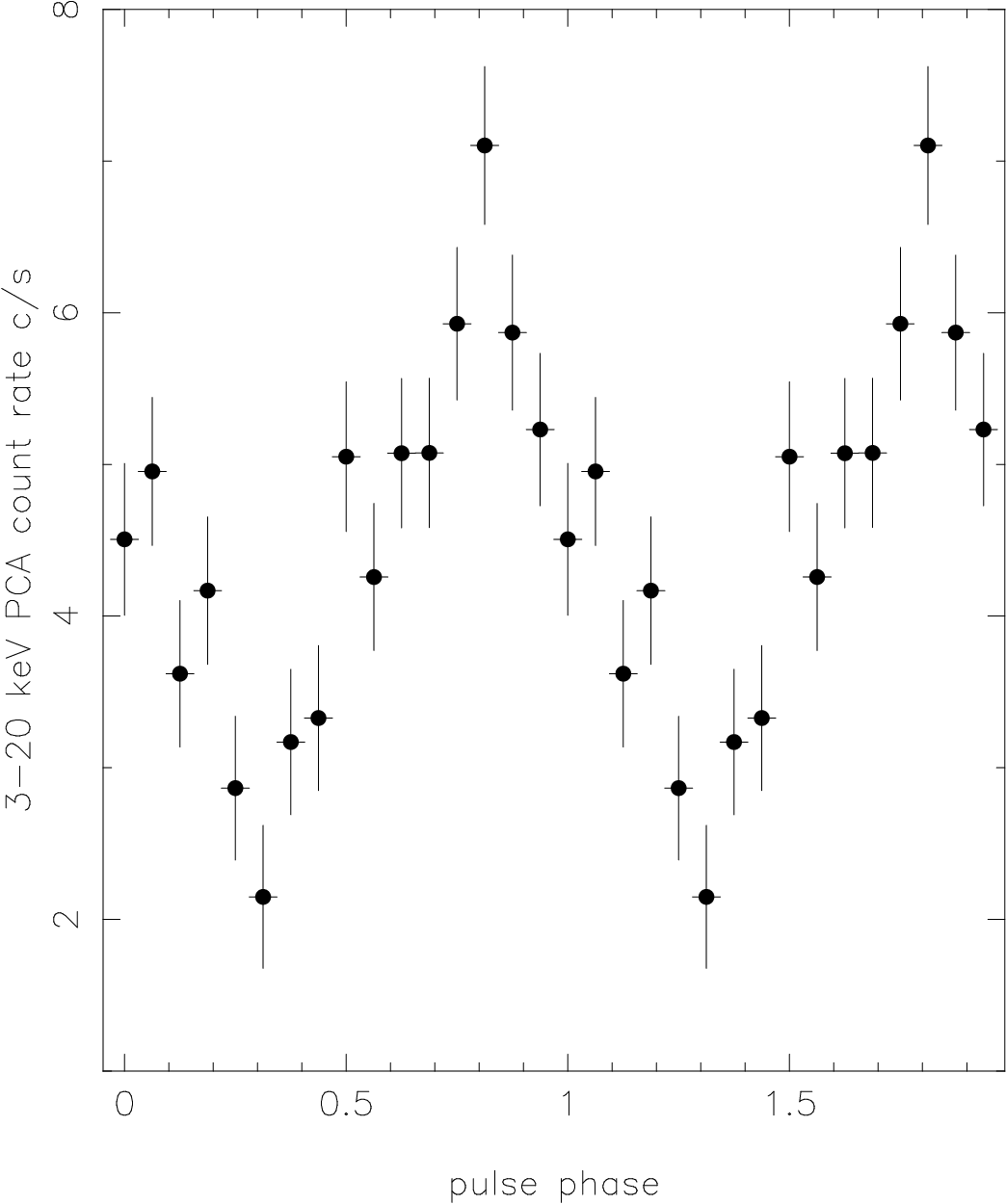}}
\end{picture}
\caption{PCA RXTE 3--20 keV background subtracted pulse profile
corresponding to the November 1998 observation.}
\label{fig:pulse}
\end{figure}

The pulse shape (see Figure~\ref{fig:pulse}) is nearly sinusoidal, as 
expected from the absence of
second or higher harmonics in the power density spectra. The amplitude of
the modulation is $\sim 2$ count s$^{-1}$ in the 3\,--\,20 keV energy range,
which implies a pulse fraction of $\sim 53$\%. Given the unknown contribution 
from Galactic Disc diffuse emission, this represents only a lower limit
to  the pulsed fraction in the signal from the source.

We have divided the November 1998 observation into two sections, 
corresponding to the peak of the pulse (phase bins 0.6 to 1.0)
and the interpulse minimum (phase bin 0.1--0.5). An absorbed power-law fit 
in the energy range 2.7\,--\,10 keV to the two spectra gave 
$\Gamma=2.9\pm0.4$, $N_{\mathrm H}=9\pm4$,
$\chi_{\mathrm r}^{2}=0.8$ (18 dof) for pulse maximum and $\Gamma=3.3\pm0.5$,
$N_{\mathrm H}=10\pm5$, $\chi_{\mathrm r}^{2}=0.9$ (18 dof) for pulse minimum.
The two values are consistent with each other within the error margins. 
The lack of spectral changes with phase requires any significant 
component of the detected flux due to diffuse emission to have a spectrum 
similar to that of the pulsar.

\subsubsection{Spectral fit}

Formally the X-ray spectra are equally 
well represented by an absorbed power-law,
blackbody and bremsstrahlung models. Table \ref{tab:models} shows the spectral
fit results. All these models gave fits of comparable quality, which means
that we are unable to distinguish meaningfully between the different
spectral models of Table \ref{tab:models}, even though the blackbody fit
is unlikely to have any physical meaning, because of very small emitting
area and the fact that it does not require any absorption (introducing 
$N_{\mathrm H}$ does not improve the fit) -- see Rutledge
et al. (1999) for a discussion of the physical inadequacy of this model
for neutron stars. The value of the hydrogen column density ($N_{\mathrm H}$),
which is consistent for the power-law and bremsstrahlung fits, is too
high to be purely interstellar. The interstellar reddening to the source
must be smaller than the measured $E(B-V) \approx 0.7$ (Steele et al. 1998),
which is the sum of interstellar and circumstellar contribution
from the disc surrounding the Be star. According to the relation by Bohlin 
et al. (1978), $E(B-V) = 0.7$ implies 
$N_{\mathrm H} =4.1\times10^{21}\:{\rm cm}^{-2}$ and therefore there must be
a substantial contribution of local material to the absorption.

From the spectral fits, we estimate the 3\,--\,20 keV X-ray
flux to be $3.5 \times 10^{33}\:{\rm erg}\,{\rm s}^{-1}$ and 
 $4.5 \times 10^{33}\:{\rm erg}\,{\rm s}^{-1}$
for the August 1998 and November 1998 observations respectively, 
assuming a distance of 2 kpc (Steele et al. 1998).
Although the values of the spectral parameters are consistent with each
other within the error margins, they all show the same trend, namely, a
harder spectral state during the 1998 August observations (lower photon
index and higher blackbody and bremsstrahlung temperatures).

\begin{table}
\caption{Results of the spectral fits. Uncertainties are given at 
$90\%$ confidence for one parameter of interest. All fits correspond to 
the energy range 3\,--\,20 keV}
\begin{center}
\begin{tabular}{lc}
\hline
\multicolumn{2}{c}{August 1998 observation} \\
\hline
\multicolumn{2}{l}{{\bf Power-law}}\\
$\Gamma$                                &  2.6$\pm$0.2 \\
$N_{\mathrm H}$ (10$^{22}$ atoms cm$^{-2}$)       &  5.6$\pm$2.2   \\
$\chi^2_{\mathrm r}$(dof)               &  1.46(43)      \\
\hline
\multicolumn{2}{l}{{\bf Blackbody}}\\
kT$_1$ (keV)                            & 1.45$\pm$0.05  \\
R (km)                                  & 0.09$\pm$0.01 \\
$\chi^2_{\mathrm r}$(dof)               &  1.73(44)      \\
\hline
\multicolumn{2}{l}{{\bf Bremsstrahlung }}\\
$kT_{\mathrm brem}$ (keV)                       &  6.4$\pm$1.3   \\
$N_{\mathrm H}$ (10$^{22}$ atoms cm$^{-2}$)       &  $2.2^{+1.8}_{-1.3}$   \\
$\chi^2_{\mathrm r}$(dof)               &  1.31(44)    \\
\hline
\hline
\multicolumn{2}{c}{November 1998 observation} \\
\hline
\multicolumn{2}{l}{{\bf Power-law}}\\
$\Gamma$                                &  3.2$\pm$0.3 \\
$N_{\mathrm H}$ (10$^{22}$ atoms cm$^{-2}$)       &  10.3$\pm$2.6   \\
$\chi^2_{\mathrm r}$(dof)               &  1.09(43)      \\
\hline
\multicolumn{2}{l}{{\bf Blackbody}}\\
$kT_1$ (keV)                            & 1.40$\pm$0.05  \\
$R$ (km)                                  & 0.11$\pm$0.01 \\
$\chi^2_{\mathrm r}$(dof)               &  1.01(44)      \\
\hline
\multicolumn{2}{l}{{\bf Bremsstrahlung }}\\
$kT_{\mathrm brem}$ (keV)                       &  4.4$\pm$0.7   \\
$N_{\mathrm H}$ (10$^{22}$ atoms cm$^{-2}$)       &  5.4$\pm$2.1   \\
$\chi^2_{\mathrm r}$(dof)               &  0.96(44)    \\
\hline
\end{tabular}
\end{center}
\label{tab:models}
\end{table}

\begin{figure}
\begin{picture}(250,280)
\put(0,0){\includegraphics{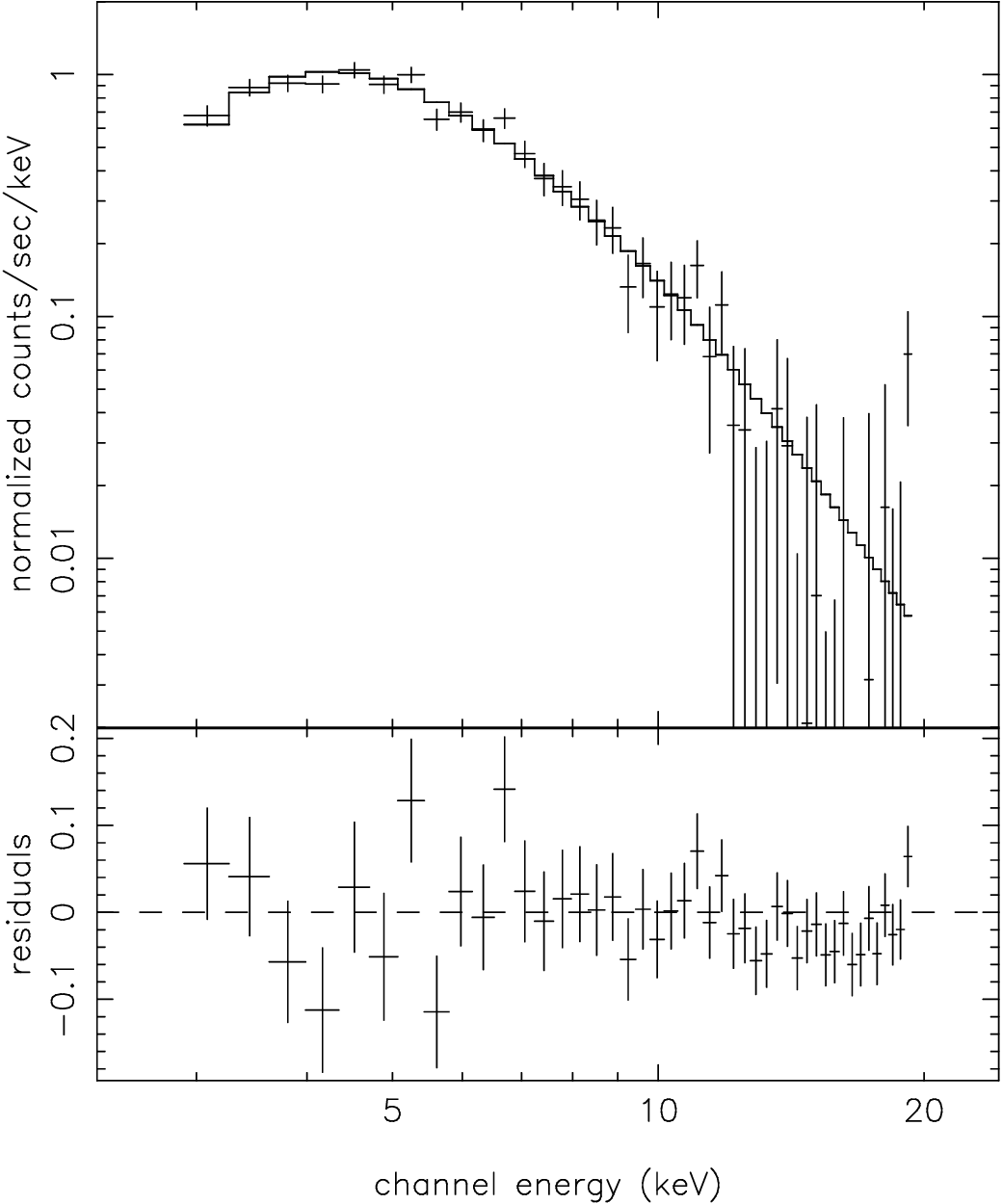}}
\end{picture}
\caption{PCA RXTE source spectrum for the November 1998 observation. 
Also shown is the best-fit power-law model and the residuals.}
\label{fig:spectrum} 
\end{figure}

\section{Results}

Our November 1998 X-ray observations represent a clear detection of 
A\,0535+26 at a time when the optical counterpart showed no evidence for 
the presence of circumstellar material. Moreover, Haigh et al. (2000) 
present spectroscopy showing that the disc was already absent as early 
as late August 1998, when our first observation was taken.
The observed luminosities in the 2\,--\,10 keV range 
($2\times10^{33}\:{\rm erg}\,{\rm s}^{-1} \la L_{{\rm x}} \la 
4.5\times10^{33}\:{\rm erg}\,{\rm s}^{-1}$) are definitely 
smaller than the quiescence luminosity observed in other occasions when the 
equatorial disc surrounding the Be star was present. For example, Motch 
et al. (1991) observed the source on several occasions 
at a level $L_{{\rm x}} \simeq 1.5\times 10^{35}\:{\rm erg}\,{\rm s}^{-1}$ 
in the 1\,--\,20 keV range (correcting 
their value to the adopted distance of 2 kpc) using {\em EXOSAT}.

The {\em EXOSAT} observation, as well as several other quiescence 
detections of Be/X-ray binaries with $P_{\mathrm s} \ga 100\:{\rm s}$,
have always been interpreted in terms of accretion on to the surface
of the neutron star from the equatorial outflow from the Be star. However, 
on this occasion we have observed the source when the disc of the Be
star had been absent for several months and at an X-ray luminosity
two orders of magnitude lower. Therefore we cannot consider as an
{\em a priori} assumption that the emission mechanism at operation is
the same.

In order to explain the difference in the luminosity by two orders of
magnitude, we can invoke accretion from a far less dense outflow or
assume that some other emission mechanism is at work. We first consider 
the possibility that the observed 
luminosity is due to accretion on to the surface of the neutron star. 
Assuming an efficiency $\eta=1$ in the conversion of gravitational 
energy into X-ray luminosity, 
$L_{{\rm x}} = 4\times10^{33}\:{\rm erg}\,{\rm s}^{-1}$
translates into an accretion rate 
$\dot{M} = 2.1\times10^{13}\:{\rm g}\,{\rm s}^{-1} = 3.3\times 10^{-13}
\:M_{\sun}\,{\rm yr}^{-1}$.

Following Stella et al. (1986), we define the corotation radius as that
at which the pulsar corotation velocity equals Keplerian velocity.
The corotation radius is given by
\begin{equation}
r_{{\rm c}} = \left( \frac{GM_{{\rm x}}P_{{\rm s}}^{2}}{4\pi^{2}}\right)^{\frac{1}{3}}
\end{equation}
where $G$ is the gravitational constant, $M_{{\rm x}}$ is the mass of 
the neutron star (assumed to be 
$1.44\:M_{\sun}$)  and $P_{{\rm s}}$ is the spin period. For A\,0535+26, 
$r_{{\rm c}} = 3.7\times10^{9}\:{\rm cm}$.

The magnetospheric radius at which the magnetic field begins to dominate
the dynamics of the inflow depends on the accretion rate and 
can be expressed as
\begin{equation}
r_{{\rm m}} = K \left(GM_{{\rm x}}\right)^{-1/7}\mu^{4/7}\dot{M}^{-2/7}
\end{equation}
where $\mu$ is the neutron star magnetic moment and $\dot{M}$
is the accretion rate. $K$ is a constant that in the case of A\,0535+26
has been determined to be $K\simeq1.0$ when an accretion disc is present
(Finger et al. 1996) and from theoretical calculations is expected to
have a similar value for wind accretion. Following
Finger et al. (1996), we will assume a magnetic dipolar field 
$9.5\times10^{12}\:{\rm G}$, resulting in $\mu= 
4.75\times10^{30}\:{\rm G}\,{\rm cm}^{3}$. 

For the accretion rate $\dot{M} = 2.1\times10^{13}\:{\rm g}\,{\rm s}^{-1}$ 
derived above,
the magnetospheric radius would be $r_{{\rm m}} = 9.3\times10^{9}\:{\rm cm}$.
Therefore $r_{{\rm m}} > r_{{\rm c}}$ and the neutron star must be in the
centrifugal inhibition regime. In order to estimate the solidity of this 
result, we point out that if the 110 keV cyclotron line detected in the
spectrum of A\,0535+26 (Kendziorra et al. 1994; Grove et al. 1995) is the
second harmonic instead of the first, as has been suggested, the magnetic
field (and magnetic moment) would be smaller by a factor 2. However, this 
would require a higher value for $K$ in order to fit the observations of a 
QPO in this system (Finger et al. 1996), leaving the value of $r_{{\rm m}}$
unaffected. An efficiency in the conversion of gravitational energy into 
radiation as low as $\eta=0.5$ will translate into a reduction in
$r_{{\rm m}}$ by only a factor $\sim 0.8$. Therefore we conclude 
that the neutron star is certain to be in the inhibited regime.

According to Corbet (1996), when the source is in the inhibition regime,
a luminosity comparable to that observed could be produced by release of
gravitational energy at the magnetospheric radius (for short, accretion 
onto the magnetosphere. However,
even assuming that the magnetosphere is  
at the corotation radius and an efficiency $\eta=1$ (i.e., best case), in
order to produce $L_{{\rm x}} = 4\times10^{33}\:{\rm erg}\,{\rm s}^{-1}$,
the accretion rate needed is
$\dot{M}_{\mathrm m} = 8.2\times10^{16}\:{\rm g}\,{\rm s}^{-1}$. With
such an accretion rate, the magnetosphere would be driven in well within
the corotation radius, and produce a luminosity of 
$L_{{\rm x}} \approx 1.5\times10^{37}\:{\rm erg}\,{\rm s}^{-1}$ by accretion
on to the surface of the neutron star. Therefore we conclude that the
observed luminosity is not due to accretion on to the magnetosphere.

Therefore we are left with the following possibilities for the origin 
of the X-ray emission:

\begin{itemize}

\item Accretion on to the neutron star through some sort of leakage through
the magnetospheric barrier. This could adopt two forms. Either directly
from the Be star outflow and only through a fairly limited region near 
the spin axis or mediated by an accretion torus, supported by the 
centrifugal barrier, with a small amount of material
managing to penetrate the magnetosphere.

\item Thermal emission from the heated core of the neutron star. Brown
et al. (1998) and Rutledge et al. (1999) have studied thermal emission
from X-ray transients in quiescence. They predict a
thermal luminosity in quiescence
\begin{equation} 
L_{\mathrm x}= 6\times10^{32}\:{\rm erg}\,{\rm s}^{-1}\times\frac{\dot{M}}
{10^{-11}\:M_{\sun}\,{\rm yr}^{-1}}
\end{equation}
where $\dot{M}$ here represents the long term average mass accretion rate. 
From the number of observed Type II outbursts in A\,0535+26, we assume
one giant outburst every 5\,--\,10 years, which translates into a long-term 
average  of $\dot{M} = 4-8\times10^{-11}\:M_{\sun}\,{\rm yr}^{-1}$. 
This would imply quiescence thermal emission in the range $L_{\mathrm x}= 
2-5\times10^{33}\:{\rm erg}\,{\rm s}^{-1}$, which is consistent
with our observations.
\end{itemize}

\section{Discussion}

It is very difficult to estimate the fraction of the signal that actually
comes from the source, though the pulsed component is evidently a lower limit 
to it. The diffuse emission from the Galactic disc is not well described
at high Galactic longitudes (for A\,0535+26, $l=181.5\degr$), but if the 
assumption by Valinia \& Marshall (1998) that its latitude distribution
should be similar to that in the Galactic Ridge can be held, then it 
should not be very strong at the position of A\,0535+26 ($b=-2.6\degr$).
In any case, the total (source + diffuse) flux detected is lower
than the average diffuse emission from the Galactic Ridge, which is 
$2.4\times10^{-11}\:{\rm erg}\,{\rm cm}^{-2}\,{\rm s}^{-1}$ in the
2\,--\,10 keV band (Valinia \& Marshall 1998).

Given that the fitted spectra are much softer than the model fits to
Galactic Ridge diffuse emission by Valinia \& Marshall (1998), which 
have photon indexes $\Gamma \sim 1.8$, and the similitude between the 
pulse-peak and pulse-minimum spectra, it seems likely that most of the
detected signal comes actually from the source. The high value of 
$N_{\mathrm H}$ obtained in all the non-thermal fits argues for the
presence of large amounts of material in the vicinity of the neutron star.
This could be caused by the pile-up of incoming material outside the 
magnetosphere. The observed spectrum
is much softer than the spectra of Be/X-ray binaries at low luminosity
during quiescence states (Motch et al. 1991 quote a photon index of
1.7 for their observations of A\,0535+26) and could favour the
thermal emission interpretation. 

Brown et al. (1998) proposed that thermal reactions deep within the 
crust of an accreting neutron star would maintain the neutron star 
core at temperatures of $\sim 10^{8}\:{\mathrm K}$, similar to that in 
a young radio pulsar. During the quiescent state of transient accretors, 
the conduction of thermal energy from the core should result in a 
detectable thermal spectrum from the neutron star atmosphere. In a 
high magnetic field pulsar, this thermal emission should be pulsed 
due to both the anisotropic surface temperature distribution caused 
by the dependence of the thermal conductivity on the magnetic field 
(Shibanov \& Yakovlev 1996), and the anisotropic radiation 
pattern from the neutron star atmosphere resulting from the magnetic 
field (Zavlin et al. 1995). 

Our blackbody fit to the A0535+26 spectrum resulted in an emission 
radius much smaller than the neutron star. Rutledge at al. (1999) show 
that fits to Hydrogen atmosphere spectral models result in larger emission 
areas and lower effective temperature than blackbody fits. The spectra may 
therefore be consistent with thermal emission from the pulsar. However, if 
this interpretation is correct, most of the emitted luminosity should be
below the energy band that we observe. In that case, the bolometric 
luminosity may well exceed that predicted 
from our estimates of the long-term average accretion rates.

Our detection of A\,0535+26 can also be used to set limits on the outflow
from the Be companion. The condition for centrifugal inhibition is
$r_{{\rm m}} \geq r_{{\rm c}}$. Therefore the minimum accretion rate
at which there is no inhibition corresponds to that at which 
$r_{{\rm m}} = r_{{\rm c}}$. Using the values above, we obtain 
an accretion rate onto the magnetosphere 
$\dot{M}_{\mathrm m} = 5.3\times10^{14}\:{\rm g}\,{\rm s}^{-1}= 
8\times10^{-12}\:M_{\sun}\,{\rm yr}^{-1}$. If the observed
luminosity is due to accretion on to the surface of the neutron star, the 
rate of mass flow into the vicinity of the neutron star is then 
constrained to be in the range $3\times10^{-13}\:M_{\sun}\,{\rm yr}^{-1}\la 
\dot{M} \la 8\times10^{-12}\:M_{\sun}\,{\rm yr}^{-1}$ (of course, if it is
due to thermal emission, only the upper limit holds). 
The lower limit represents a small fraction of the mass lost from the 
Be star (which should be $\ga 10^{-11}\:M_{\sun}\,{\rm yr}^{-1}$), but
the upper limit is close to the mass-loss values derived in the
decretion disc model (Okazaki 1997b; Porter 1999).  This value is also close
to the accretion rate needed to sustain the quiescence luminosity
observed by Motch et al. (1991), which is 
$\dot{M} \sim 10^{-11}\:M_{\sun}\,{\rm yr}^{-1}$. Such an accretion
rate would represent a substantial fraction of the stellar 
mass-loss, though still only a small fraction of the disc mass (estimated
to be $ 10^{-9}\,-\,10^{-10}\:M_{\sun}$). As has been calculated above,
most of the long-term accretion rate is due to the giant outbursts, where
a substantial fraction of the Be disc mass must be accreted.

From all the above, it is clear that the amount of material reaching 
the vicinity of the neutron
star during the disc-less phase of the companion is smaller than during
previous quiescence states. This cannot be due to an orbital effect because
Motch et al. (1991) observed the source at different orbital times and
also because our two observations took place close to periastron (orbital
phases $\phi \simeq 0$ for the August observation and $\phi \simeq 0.8$ for 
the November observation, according to the ephemeris of Finger et al. 
1996). 

Existing evidence seems to argue against the existence of a persistent 
accretion disc surrounding the neutron star in A\,0535+26. The statistical 
analysis of Clark et al. (1999) showed that there is no
significant contribution from an accretion disc to the optical/infrared
luminosity of A\,0535+26. This does not rule out the presence of an accretion
disc (which could, for example, be too small to radiate a significant 
flux in comparison to the Be circumstellar disc). It is believed that 
A\,0535+26 does indeed form an accretion disc around
the neutron during Type II outbursts, since very fast spin-up and 
quasi-periodic oscillations have been observed (Finger et al. 1996).
However, the lack of spin-up during Type I outbursts led Motch et al. (1991)
to conclude that no persistent accretion disc was present.
In contrast, the Be/X-ray binary 2S\,1845$-$024 shows large spin-up 
during every Type I outburst  (Finger et al. 1999). If no accretion disc
is present, the reduced amount
of material reaching the neutron star must be directly due to a change in the
parameters of the outflow from the Be star. Within the framework of modern
models for Be star discs, considering very subsonic outflow velocities, 
such a change can only be due to a lower outflow density. Unfortunately,
since the details of the magnetic inhibition process are poorly understood,
we can only constrain the mass rate reaching the neutron star to be
below that corresponding to the transition at which inhibition occurs,
which is very close to the rate deduced from previous quiescence 
observations in which the source was not in the inhibition regime.

\section{Conclusions}

A\,0535+26 was active at a time when the optical counterpart V725 Tau
showed no evidence for a circumstellar disc. The luminosity was two
orders of magnitude lower than in previous quiescence detections and
accretion was centrifugally inhibited. Given that the observed luminosity
cannot be due to accretion onto the magnetosphere, we are observing
either some material leaking through the magnetosphere or thermal 
emission from the heated core of the neutron star. In any case, this
detection represents a state of an accreting X-ray pulsar that had not
been observed before.

Further observations of Be/X-ray binaries in a similar state (when their
companions have lost their discs and very little material can reach the
vicinity of the neutron star) are needed. Observations with {\em Chandra} 
or {\em XMM}, which combine much higher sensitivities with more adequate
energy ranges, could
determine whether the observed spectrum is compatible with thermal emission
models.

\section*{Acknowledgements}

We thank Jean Swank and the {\em RossiXTE} team for granting a Target of
Opportunity observation and carrying it out on very short notice. Simon
Clark is thanked for his help in preparing the proposal. The WHT is
operated on the  island of La Palma by the Royal Greenwich Observatory in 
the Spanish  Observatorio del Roque de Los Muchachos of the Instituto de
Astrof\'{\i}sica de Canarias. The observations were taken as part of the
ING service observing programme. This research has made use of data 
obtained through the High Energy Astrophysics Science Archive Research 
Center Online Service, provided by the NASA/Goddard Space Flight Center.

IN is supported by an ESA external fellowship. P.~Reig acknowledges partial 
support via the European Union Training and Mobility of Researchers Network
Grant ERBFMRX/CT98/0195.

\end{document}